\documentclass[twocolumn]{aastex61}

\usepackage{amsmath}

\newcommand{\beq}{\begin{equation}}
\newcommand{\eeq}{\end{equation}}

\newcommand\reye{\mathrm{Re}}
\newcommand\reym{\mathrm{Rm}}
\newcommand{\Pm}{\mathrm{Pm}}
\newcommand{\Co}{\mathrm{Co}}

\newcommand{\uphi}{\ensuremath{u_\phi}}
\newcommand{\rhat}{\ensuremath{\mathbf{\hat{r}}}}
\newcommand{\phihat}{\ensuremath{\mathbf{\hat{\phi}}}}
\newcommand{\zhat}{\ensuremath{\mathbf{\hat{z}}}}

\shorttitle{The weakly nonlinear magnetorotational instability}
\shortauthors{Clark \& Oishi}

\begin{document}

\title{The Weakly Nonlinear Magnetorotational Instability in a Global, Cylindrical Taylor-Couette Flow}

\correspondingauthor{S. E. Clark}
\email{seclark@astro.columbia.edu}

\author[0000-0002-7633-3376]{S. E. Clark}
\affil{Department of Astronomy, Columbia University, New York, NY}

\author{Jeffrey S. Oishi}
\affiliation{Department of Physics and Astronomy, Bates College, Lewiston, ME}
\affiliation{Department of Astrophysics, American Museum of Natural History, New York, NY}
\nocollaboration

\begin{abstract}
We conduct a global, weakly nonlinear analysis of the magnetorotational instability (MRI) in a Taylor-Couette flow. This is a multiscale perturbative treatment of the nonideal, axisymmetric MRI near threshold, subject to realistic radial boundary conditions and cylindrical geometry. We analyze both the standard MRI, initialized by a constant vertical background magnetic field, and the helical MRI, with an azimuthal background field component. This is the first weakly nonlinear analysis of the MRI in a global Taylor-Couette geometry, as well as the first weakly nonlinear analysis of the helical MRI. We find that the evolution of the amplitude of the standard MRI is described by a real Ginzburg-Landau equation (GLE), while the amplitude of the helical MRI takes the form of a complex GLE. This suggests that the saturated state of the helical MRI may itself be unstable on long spatial and temporal scales.  
\end{abstract}
\section{Introduction}

The magnetorotational instability (MRI) is believed to drive angular momentum transport in astrophysical disks. The MRI is a local instability excited by weak magnetic fields in differentially rotating fluids, and since first applied to an astrophysical context \citep{Balbus:1991vs} it has been invoked to explain accretion in protoplanetary disks \citep{Armitage:2010} and disks around black holes \citep{Blaes:2014}, as well as jet and wind launching \citep{Lesur:2013}, anisotropic turbulence \citep{Murphy:2015tn}, and dynamo generation \citep{Brandenburg:1995, Vishniac:2009il}. 

\begin{table*}[ht]
\normalsize
\caption{Fiducial parameters for MRI runs} \label{table:parameters}
\centering
\begin{tabular}{rlrrrrrrr}
  \hline
 & $\xi$ & $\Pm$ & $\Co$ & $\Omega_2/\Omega_1$ & $R_1/R_2$ & radial magnetic b.c. \\ 
  \hline\hline
Standard MRI & 0 & 1.6E-6 & 4.85E-2 & 0.121 & 0.33 & conducting \\ 
Helical MRI & 4 & 1E-6 & 118 & 0.27 & 0.5 & insulating\\ 
   \hline
\end{tabular}
\end{table*} 

The diversity of astrophysical systems which may be MRI unstable yields an enormous parameter space to be explored. In protoplanetary disks, for example, the behavior and evolution of the MRI -- and even its very existence -- may change drastically depending on the properties of the magnetic field, the disk composition, disk geometry, and so forth. Multiphysics numerical simulations of such systems are currently an area of intense focus, enabling the study of nonideal MHD effects, disk stratification, nonequilibrium chemistry, and other complex physics that does not lend itself easily to analytic study \citep[e.g.][among many others]{Fleming:2003fs,Bai:2011cm, Flock:2013,Suzuki:2014vh}. Still, computational costs inevitably constrain numerical approaches. MRI saturation is a complicated nonlinear problem which may depend on the assumptions and approximations adopted by simulations in nonobvious ways. For example, the magnetic Prandtl number $\Pm = \nu/\eta \sim 10^{-8}$ in protoplanetary disks \citep[e.g.][]{Oishi:2011ei} and $\sim 10^{-6}$ in liquid metal experiments \citep[e.g.][]{Goodman:2002ix}. Such extreme ratios of viscosity to resistivity far exceed current computational resources. However, we can construct asymptotic approximations valid for $\Pm \ll 1$ using analytic methods. 

Analytic methods can also play a powerful role in elucidating the mechanisms responsible for MRI saturation. For instance, analytical approaches have revealed the mechanism that likely governs saturation in the ``shearing box'' approximation. The shearing box is an oft-invoked local approximation in which a section of a disk is represented by solving the MHD equations in a rotating, Cartesian box with a linearized background shear, subject to shear periodic boundary conditions in the radial direction. The shearing box is a convenient computational framework allowing extreme resolution for local MRI studies and has been extended to include vertical stratification and a wide variety of diffusive effects. 

However, while the MRI is a local instability, there are a number of important problems that require a global treatment. Perhaps most importantly, linear evolution in the shearing box is dominated by channel modes, particularly when a net vertical magnetic field threads the box. These linear modes are exact solutions to the \textit{nonlinear} local MRI equations. The shearing box MRI system avoids runaway growth by a secondary instability of the channel modes themselves \citep{Goodman:1994ul, Pessah:2010ic}. The growth of parasitic modes provides a saturation avenue for channel mode-dominated flows, yet this is unlikely to be the dominant saturation mechanism in laboratory experiments or astrophysical disks, as channel modes are artificially over-represented in the shearing box \citep{Latter:2015}. Thus while the shearing box may accurately approximate many features of the global MRI, the saturation mechanism may not be among them. In \citet[][hereafter Paper I]{Clark:2016} we find that the fastest-growing MRI mode in the shearing box is not a channel mode when the effects of ambipolar diffusion are formally included. It is thus important to ask whether the symmetries that give rise to the weakly nonlinear saturation in the local geometry are also manifested in the global flow.

In this paper, we develop a weakly nonlinear, global theory for the MRI in a Taylor-Couette (TC) geometry. This system precludes channel modes, allowing us to develop an understanding of MRI saturation in their absence. A number of saturation mechanisms have been proposed for the MRI which do not rely on channel modes dominating the flow. The MRI feeds off of the free energy from differential rotation, and so a modification of the background shear may cause saturation \citep{Knobloch:2005ba, Umurhan:2007hs}. The MRI may transfer its free energy into the magnetic field, and saturate when the field is too strong to be susceptible to the MRI \citep{Ebrahimi:2009ey}. The MRI may saturate differently depending on the particular parameter regime under investigation, and so our challenge is not only in identifying possible saturation mechanisms, but in understanding how and when each applies in different astrophysical environments. 

\begin{figure}
\centering
\includegraphics[width=\columnwidth]{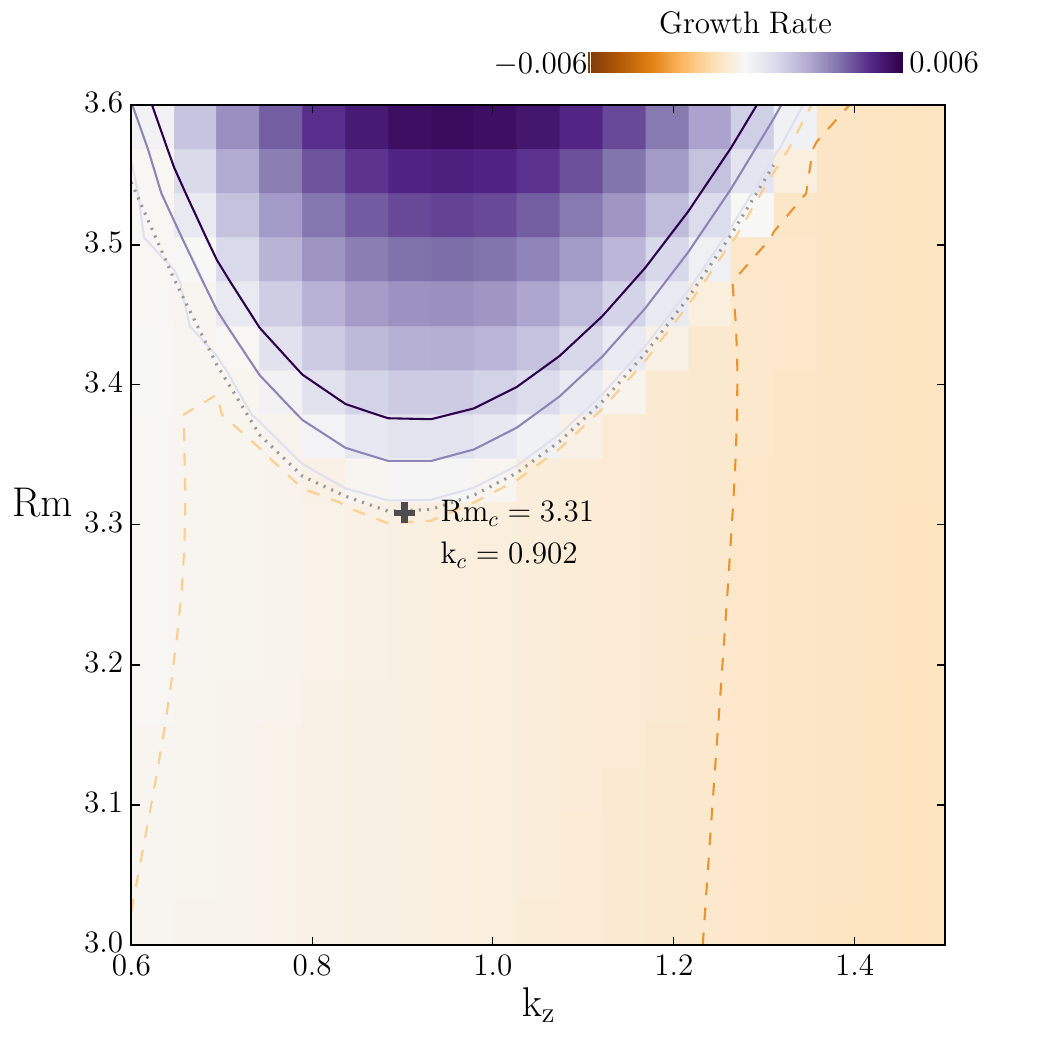}
\caption{Growth rates in the ($\reym, k_z$) plane. Color map shows growth rate found by solving the linear eigenvalue problem for each ($\reym, k_z$) in the grid. The eigenvalue problem was solved for the widegap parameters listed in Table \ref{table:parameters}. Overlaid contours show growth rates at [-8E-4, -1.3E-4, 1.3E-4, 8E-4, 1.5E-3], where dashed contours represent negative values. The gray dotted line shows the interpolated marginal stability curve. The critical parameters $\reym_c = 3.31$ and $k_c = 0.902$ correspond to the smallest parameter values that yield a zero growth rate.}\label{fig:growth_rates}
\end{figure}

Our investigation is astrophysically motivated, but we also intend our theory to be relevant to laboratory experiments. Several experimental efforts are attempting to observe the MRI in the laboratory, which will allow the study of a crucial astrophysical phenomenon in a controlled setting. Unfortunately, detection of the MRI has so far proven elusive. \citet{Sisan:2004ig} claimed to detect the MRI in a spherical Couette flow, but most likely detected unrelated MHD instabilities instead \citep{Hollerbach:2009ig,Gissinger:2011td}. Most relevant to our work is the Princeton Plasma Physics Laboratory (PPPL) MRI Experiment, a liquid gallium TC flow with an axial magnetic field \citep{Ji:2001kd}. There has been a significant amount of theoretical work designed to complement the Princeton MRI experiment involving direct numerical simulation of the experimental conditions, much of it focused on the specific challenges in identifying MRI signatures despite spurious, apparatus-driven flows \citep[e.g.][]{Gissinger:2012gc}. The vertical endcaps on a laboratory MRI apparatus drive meridional flows which both inhibit the excitement of MRI and obscure its detection. The Princeton MRI experiment employs split, independently rotating endcaps to mitigate these flows \citep{Schartman:2009df}. Our work assumes an infinite vertical domain, an idealization that is theoretically expedient but experimentally impractical. Such an approach changes the symmetry properties of the solution significantly, and in the much better studied hydrodynamic case this leads to significant differences even for TC devices with very large aspect ratios \citep{Lopez:2005}. Nevertheless, this study represents a first step in understanding the saturation of global, MRI unstable TC flow without the additional complication of vertical endcap effects.

\begin{figure*}
\centering
\includegraphics[width=\textwidth]{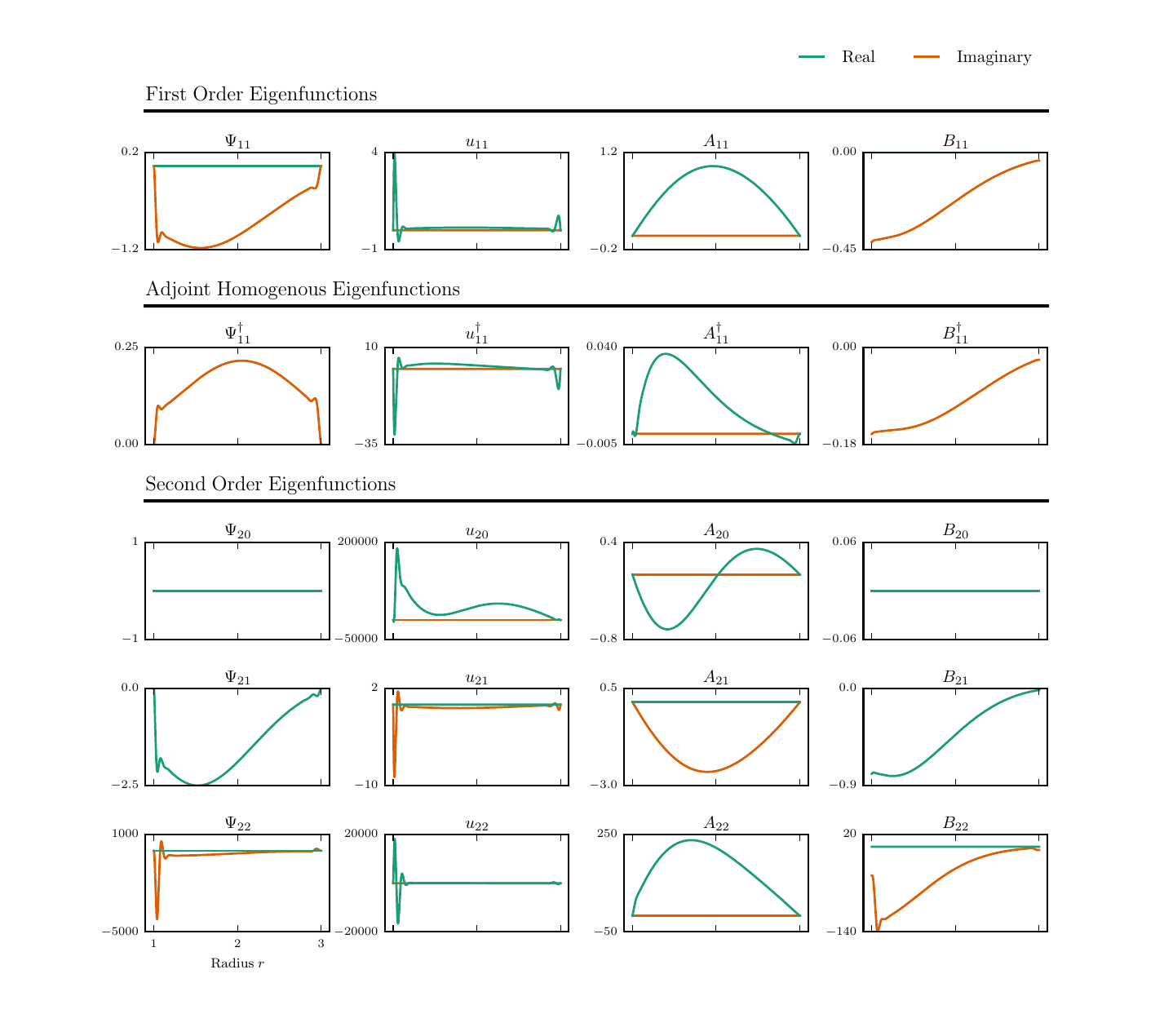}
\caption{Eigenfunctions of the first order equations, first order adjoint homogenous equations, and second order equations. We use our fiducial parameters for the standard MRI ($\xi = 0$). Eigenfunctions are solved on a 512-element grid of Chebyshev polynomials. First-order eigenfunctions are normalized such that $A_{11}(r_0) = 1$. Adjoint homogenous eigenfunctions are normalized such that $\langle V_{11}^\dagger \cdot \mathcal{D} V_{11} \rangle = 1$.}\label{fig:linear_eigenfunctions}
\end{figure*}

Our radial treatment includes the curvature of the flow in a cylindrical apparatus. Many investigations of the MRI use the ``narrow gap'' approximation (the shearing box is a narrow gap without boundary walls), in which the radial extent of the fluid channel is taken to be much smaller than the radius of curvature. That is, for a center channel radius $r_0$ bounded by inner and outer radii $r_1$ and $r_2$, respectively, the narrow gap approximation applies when $r_0 \gg (r_2 - r_1)$. The narrow gap approximation simplifies the MRI equations by excluding curvature terms, because the flow through a narrow gap can be taken to be approximately linear in $\phi$, i.e. Cartesian. Previous investigations into the weakly nonlinear behavior of the MRI have used this narrow gap approximation \citep{Umurhan:2007dz,Umurhan:2007hs,Clark:2016}. Building on our work in Paper I, here we undertake the first (to our knowledge) weakly nonlinear analysis of the MRI in the wide gap regime, where the channel width may be comparable to or larger than its distance from the center of rotation.

Because we include curvature terms, our treatment also allows us to study the helical magnetorotational instability. The helical MRI is an overstability in which the background magnetic field is helical, $\mathbf{B} = B_0 (\xi r/r_0 \mathbf{\hat{\phi}} + \mathbf{\hat{z}})$ \citet{Hollerbach:2005tr}. The helical MRI currently occupies a special place in the MRI puzzle. The helical MRI has been proposed as a method of awakening angular momentum transport in the ``dead zones" of protoplanetary disks where the $\reym$ becomes very small. However the rotation profiles needed to excite helical MRI may be steeper than Keplerian, depending on the boundary conditions, and so its role in astrophysical disks is currently a matter of debate \citep{Liu:2006,Rudiger:2007,Kirillov:2013}. Regardless of its astrophysical role, the helical MRI is significantly easier to excite in a laboratory setting than the standard MRI, and has already been detected by the Potsdam Rossendorf Magnetic Instability Experiment \citep[PROMISE;][]{Stefani:2006iv,Stefani:2009hp}.

In this work we explore the behavior of the viscous, dissipative MRI in a cylindrical geometry close to threshold, making explicit comparisons to the standard MRI behavior in the thin-gap regime. We investigate both the standard MRI, in which the background magnetic field is purely axial, as well as the helical MRI. In section~\ref{sec:wide_gap}, we lay out the basic mathematical framework of the problem. In section~\ref{sec:wnl_analysis}, we introduce the method of multiple scales we use to construct our theory. In section~\ref{sec:results} we describe the basic results, and in section~\ref{sec:discussion} we place them in the context of previous work on other instabilities, discuss their relevance to experiments, and reiterate our final conclusions.

\begin{figure*}
\centering
\includegraphics[width=\textwidth]{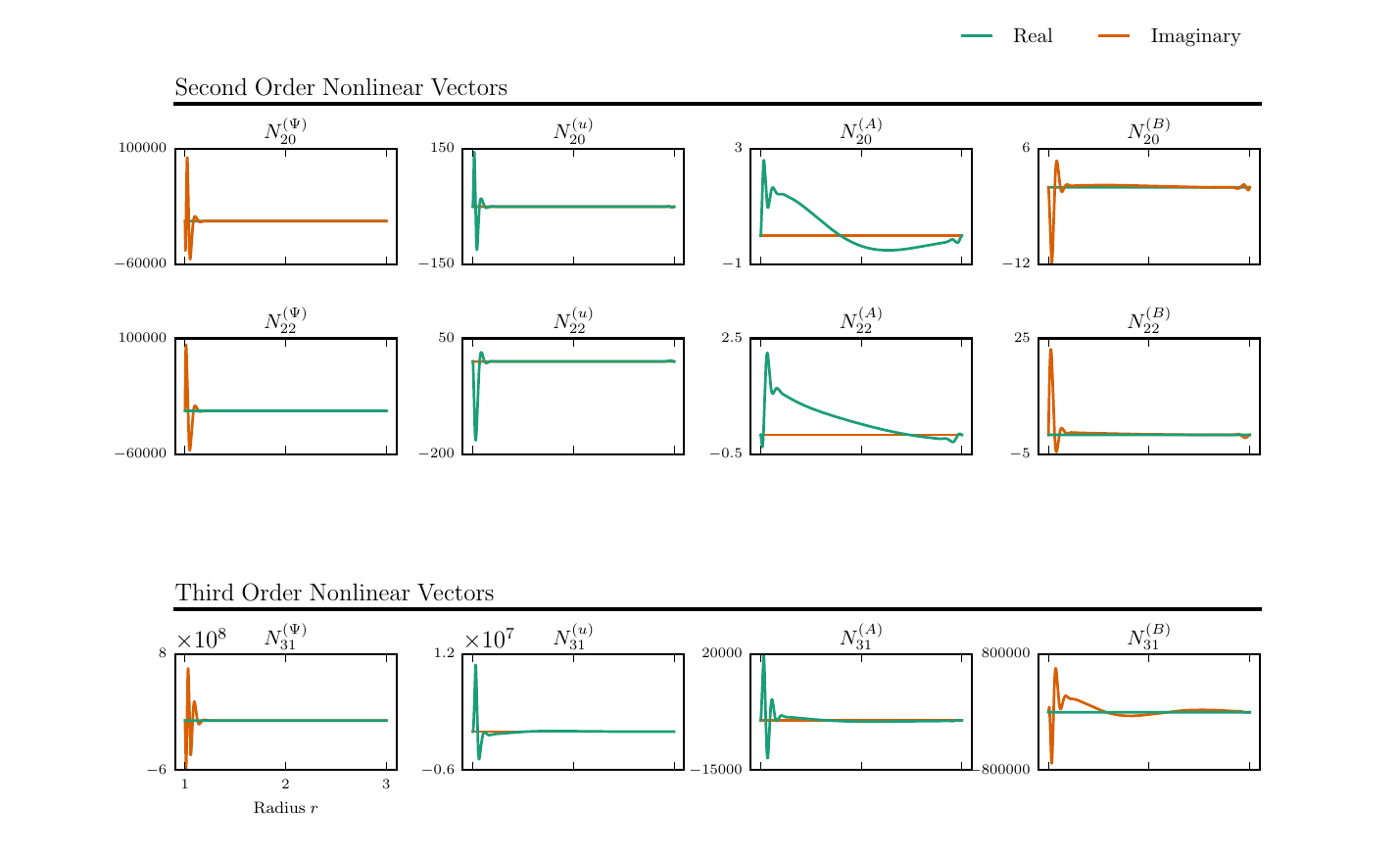}
\caption{Nonlinear terms $N_2$ and $N_3$ for our fiducial standard MRI parameters. These are nonlinear combinations of lower-order eigenfunctions. At second order ($N_2$) the most unstable linear MRI mode interacts with itself and its complex conjugate. At third order ($N_3$) the first and second order MRI modes interact with each other.}\label{fig:N2_N3}
\end{figure*}

\section{Basic Framework}
\label{sec:wide_gap}
The basic equations solved are the momentum and induction equations,

\beq\label{momentum}
\partial_t \mathbf{u} + \mathbf{u} \cdot \nabla \mathbf{u} = -\frac{1}{\rho}\nabla P - \nabla\Phi + \frac{1}{c\rho} \left(\mathbf{J}\times\mathbf{B}\right) + \nu\nabla^2 \mathbf{u} 
\eeq

and

\beq\label{induction}
\partial_t \mathbf{B} = \nabla \times \left(\mathbf{u} \times \mathbf{B}\right) + \eta\nabla^2\mathbf{B},
\eeq

where $P$ is the gas pressure, $\nu$ is the kinematic viscosity, $\eta$ is the microscopic diffusivity, $\nabla\Phi$ is the gravitational force per unit mass, and the current density is $\mathbf{J} = c \nabla\times\mathbf{B}/4\pi$. We solve these equations subject to the incompressible fluid and solenoidal magnetic field constraints,

\beq
\label{eq:incompressibility}
\nabla \cdot \mathbf{u} = 0
\eeq

and 

\beq
\label{eq:solenoid}
\nabla \cdot \mathbf{B} = 0.
\eeq

We perturb these equations axisymmetrically in a cylindrical $(r, \phi, z)$ geometry, i.e. $\mathbf{u} = \mathbf{u_0} + \mathbf{u_1}$ and $\mathbf{B} = \mathbf{B_0} + \mathbf{B_1}$, where $\mathbf{u_0}$ and $\mathbf{B_0}$ are defined below. We define a Stokes stream function $\Psi$ such that 

\beq
  \label{eq:stokes}
  \mathbf{u_1} \, = \, \left[\begin{matrix}
\frac{1}{r} \partial_z \Psi\ \rhat\\
\uphi \ \phihat\\
-\frac{1}{r} \partial_r \Psi\ \zhat\\
\end{matrix}\right],
\eeq

and the magnetic vector potential $A$ is

\beq
  \label{eq:fluxfunc}
  \mathbf{B_1} \, = \, \left[\begin{matrix}
\frac{1}{r} \partial_z A\ \rhat\\
B_{\phi} \ \phihat\\
-\frac{1}{r} \partial_r A\ \zhat\\
\end{matrix}\right].
\eeq

These definitions automatically satisfy Equations \ref{eq:incompressibility} and \ref{eq:solenoid} for axisymmetric disturbances. We note that in the linearized equations, streamfunctions of the form $u_x = \partial_z \Psi$, $u_z = -(\partial_r + \frac{1}{r}) \Psi$, and the corresponding definitions of the magnetic vector potential, are convenient choices, but we define Equations \ref{eq:stokes} and \ref{eq:fluxfunc} for this nonlinear investigation because of the incommutability of $\partial_r$ and $\partial_r + \frac{1}{r}$. 

The astrophysical magnetorotational instability operates in accretion disks and in stellar interiors, environments where fluid rotation is strongly regulated by gravity. In accretion disks, differential rotation is imposed gravitationally by a central body, so the rotation profile is forced to be Keplerian. Clearly a gravitationally enforced Keplerian flow is inaccessible to laboratory study, so differential rotation is created by rotating an inner cylinder faster than an outer cylinder (a TC setup). For a nonideal fluid subject to no-slip boundary conditions, the base flow is

\begin{equation}
  \label{eq:couette_flow}
  \Omega(r) = c_1 + \frac{c_2}{r^2},
\end{equation}
where $c_1 = (\Omega_2 r^2_2 - \Omega_1 r^2_1)/(r^2_2 - r^2_1)$, $c_2 = r^2_1 r^2_2 (\Omega_1 - \Omega_2)/(r^2_2 - r^2_1)$, and $\Omega_1$ and $\Omega_2$ are the rotation rates at the inner and outer cylinder radii, respectively. In the laboratory, $r_1$ and $r_2$ are typically fixed by experimental design. However $\Omega_1$ and $\Omega_2$ may be chosen such that the flow in the center of the channel is approximately Keplerian. Defining a shear parameter $q$, we see that for Couette flow,
\begin{equation}
  \label{eq:couette_q}
  q(r) \equiv -\frac{d \ln \Omega}{d \ln r} = \frac{2 c_2}{c_1 r^2 + c_2}.
\end{equation}

Thus through judicious choice of cylinder rotation rates, one can set $q(r_0) = 3/2$, for quasi-Keplerian flow. Note that the narrow gap approximation imposes a linear shear (constant $q$), and so the interaction of fluid perturbations with the base velocity profile differs significantly from the case considered here. Our base velocity is 

\beq
\label{eq:baseu0}
\mathbf{u_0} = r\Omega(r) \phihat.
\eeq

We initialize a magnetic field 

\beq
\label{eq:baseB0}
\mathbf{B_0} = B_0 \zhat + B_0 \xi \frac{r_0}{r} \phihat,
\eeq

so that the base magnetic field is axial when $\xi = 0$ and otherwise helical. 

In this work we will focus our findings on two fiducial parameter sets, one for the standard MRI where $\xi = 0$ and one for the helical MRI. We choose the standard MRI parameters to be comparable to the case considered in \citet{Goodman:2002ix}. The helical MRI parameters were chosen to be comparable to \citet{Hollerbach:2005tr}. Our fiducial parameters are described in Table \ref{table:parameters}.

Our perturbed system is 
\onecolumngrid
\beq\label{eq:Psi_perturbed}
\frac{1}{r}\partial_t (\nabla^2 \Psi - \frac{2}{r} \partial_r \Psi) - \Co \frac{1}{r}B_0 \partial_z (\nabla^2 A - \frac{2}{r} \partial_r A) - \frac{2}{r}u_0 \partial_z u_\phi + \Co \frac{2}{r^2}B_0 \xi \partial_z B_\phi - \frac{1}{\reye} \left[ \nabla^2 (\frac{1}{r} \nabla^2 \Psi - \frac{2}{r^2}\partial_r\Psi) - \frac{1}{r^3} \nabla^2 \Psi + \frac{2}{r^4}\partial_r\Psi\right] = N^{(\Psi)}
\eeq

\beq \label{eq:uphi_perturbed}
\partial_t \uphi + \frac{1}{r^2} u_0 \partial_z \Psi + \frac{1}{r} \partial_r u_0 \partial_z \Psi - \Co B_0 \partial_z B_\phi - \frac{1}{\reye} ( \nabla^2 \uphi - \frac{1}{r^2} \uphi ) = N^{(u)}
\eeq

\beq
\partial_t A - B_0 \partial_z \Psi - \frac{1}{\reym} (\nabla^2 A - \frac{2}{r} \partial_r A) = N^{(A)}  \label{eq:A_perturbed}
\eeq

\beq  
\label{eq:Bphi_perturbed}
\partial_t B_\phi + \frac{1}{r^2} u_0 \partial_z A - B_0 \partial_z u_\phi - \frac{1}{r} \partial_r u_0 \partial_z A - \frac{2}{r^3} B_0 \xi \partial_z \Psi - \frac{1}{\reym} (\nabla^2 B_\phi - \frac{1}{r^2} B_\phi ) = N^{(B)}
\eeq

The righthand side of the equations contain the nonlinear terms

\beq
N^{(\Psi)} = - J(\Psi, \frac{1}{r^2} ( \nabla^2 \Psi - \frac{2}{r} \partial_r\Psi) ) + \Co J(A, \frac{1}{r^2} ( \nabla^2 A - \frac{2}{r} \partial_rA) ) - \Co \frac{2}{r}B_\phi \partial_z B_\phi  + \frac{2}{r} u_\phi \partial_z u_\phi 
\eeq
\beq
N^{(u)} = \Co \frac{1}{r} J(A, B_\phi) - \frac{1}{r} J(\Psi, \uphi) + \Co \frac{1}{r^2} B_\phi \partial_z A - \frac{1}{r^2} \uphi \partial_z \Psi 
\eeq
\beq
N^{(A)} = \frac{1}{r} J(A, \psi)
\eeq
\beq
N^{(B)} = \frac{1}{r} J(A, \uphi) + \frac{1}{r} J(B_\phi, \psi) + \frac{1}{r^2} B_\phi \partial_z \psi - \frac{1}{r^2} \uphi \partial_z A 
\label{eq:nonlinear_B}
\eeq
\twocolumngrid

where J is the Jacobian $J(f, g) \equiv \partial_z f \partial_r g - \partial_r f \partial_z g$. Note that in the above, $\nabla^2 f \equiv \partial_r^2 f + \partial_z^2 f + \frac{1}{r} \partial_r f$. Equations \ref{eq:Psi_perturbed} - \ref{eq:nonlinear_B} are nondimensionalized by inner cylinder quantities: lengths have been scaled by $r_1$, velocities by $r_1 \Omega_1$, and densities by $\rho_0$, where $\rho_0$ is the constant density. Magnetic fields are scaled by $B_0$, the constant strength of the initial background field; where $B_0$ appears in the above it is formally unity. $\Omega_1 = \Omega(r_1)$ is the rotation rate of the inner cylinder. We introduce the Reynolds number $\reye = \Omega_1 r_1^2/\nu$, the magnetic Reynolds number $\reym = \Omega_1 r_1^2 / \eta$, and a plasma beta parameter $\Co = 2 B_0^2 / \Omega_1^2 r_1^2 \rho_0$. Note that if we define the dimensional cylindrical coordinate $r = r_1(1 + \delta x)$, we recover the narrow gap approximation of the system in the limit $\delta \rightarrow 0$.

We solve the standard MRI system subject to the same boundary conditions used in \citet{Goodman:2002ix}. These are periodic vertical boundary conditions and no-slip, perfectly conducting radial boundary conditions, namely

\beq
\Psi = \partial_r \Psi = u = A = \partial_r (r B) = 0
\eeq

at $r = r_1, r_2$. To the helical MRI system we apply insulating boundary conditions as used in \citet{Hollerbach:2005tr}:

\beq
\partial_r A = k \frac{I_0 (k r)}{I_1 (k r)} A \, \, \mathrm{at} \, \, r = r_1
\eeq

\beq
\partial_r A = - k \frac{K_0 (k r)}{K_1 (k r)} A \, \, \mathrm{at} \, \, r = r_2
\eeq

and $B = 0$ at $r = r_1, r_2$ \citep[see][]{Willis:2002bh}. Here, $I_n$ and $K_n$ are the modified Bessel functions of the first and second kind, respectively.

We note that Equations \ref{eq:Psi_perturbed} - \ref{eq:Bphi_perturbed} are written in a nonstandard form, with the nonlinear terms on the righthand side. This choice has a practical motivation. As detailed in \S\ref{sec:wnl_analysis}, we expand these equations in a perturbation series and solve them order by order using a pseudospectral code. The code solves partial differential equations of the form $\mathrm{M} \partial_t \mathbf{V} + \mathrm{L} \mathbf{V} = \mathbf{F}$, where $\mathrm{M}$ and $\mathrm{L}$ are matrices and $\mathbf{F}$ is a vector containing any nonhomogenous terms. The nonlinear terms in our perturbation analysis become nonhomogenous term inputs to the solver.

\begin{figure*}
\centering
\includegraphics[width=\textwidth]{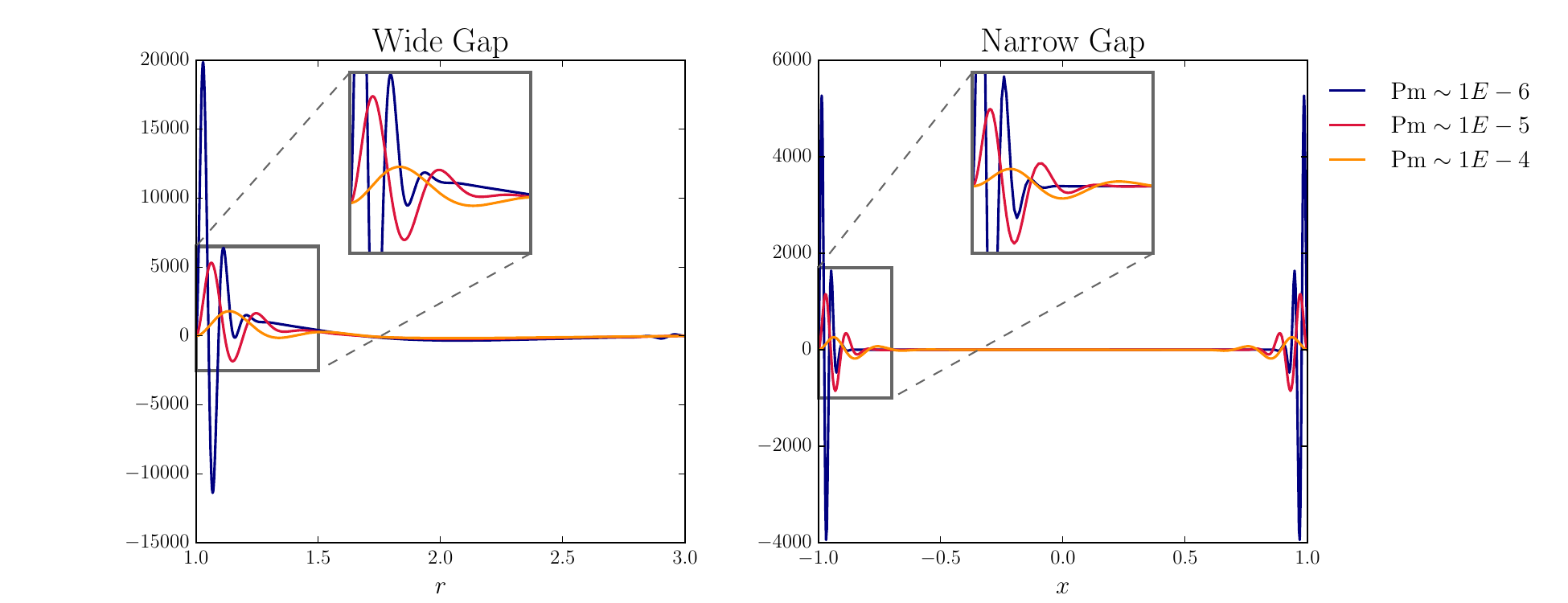}
\caption{Nonlinear term $N_{31}^{(A)}$ for the wide gap (left) and narrow gap (right) standard MRI, where the wide gap is the TC flow considered in this work. Terms shown span three orders of magnitude in $\Pm$. The wide gap vectors represent runs using the parameters in Table \ref{table:parameters} and $\Pm = 1.6E-4, 1.6E-5, 1.6E-6$. The narrow gap MRI runs use the fiducial parameters in \citet{Umurhan:2007hs}, with $\Pm = 1E-4, 1E-5, 1E-6$. Inlaid plots show zoomed-in views of boundary layers at the inner boundary. The wide gap case displays dramatic boundary layers only at the inner boundary, but boundary layers in the thin gap approximation are symmetric about the origin because MRI modes in the narrow gap approximation are eigenstates of parity.}\label{fig:thin_wide_comparison} 
\end{figure*}

\section{Weakly Nonlinear Perturbation Analysis}\label{sec:wnl_analysis}
\label{sec:Perturbation}

\begin{figure}
\centering
\includegraphics[width=\columnwidth]{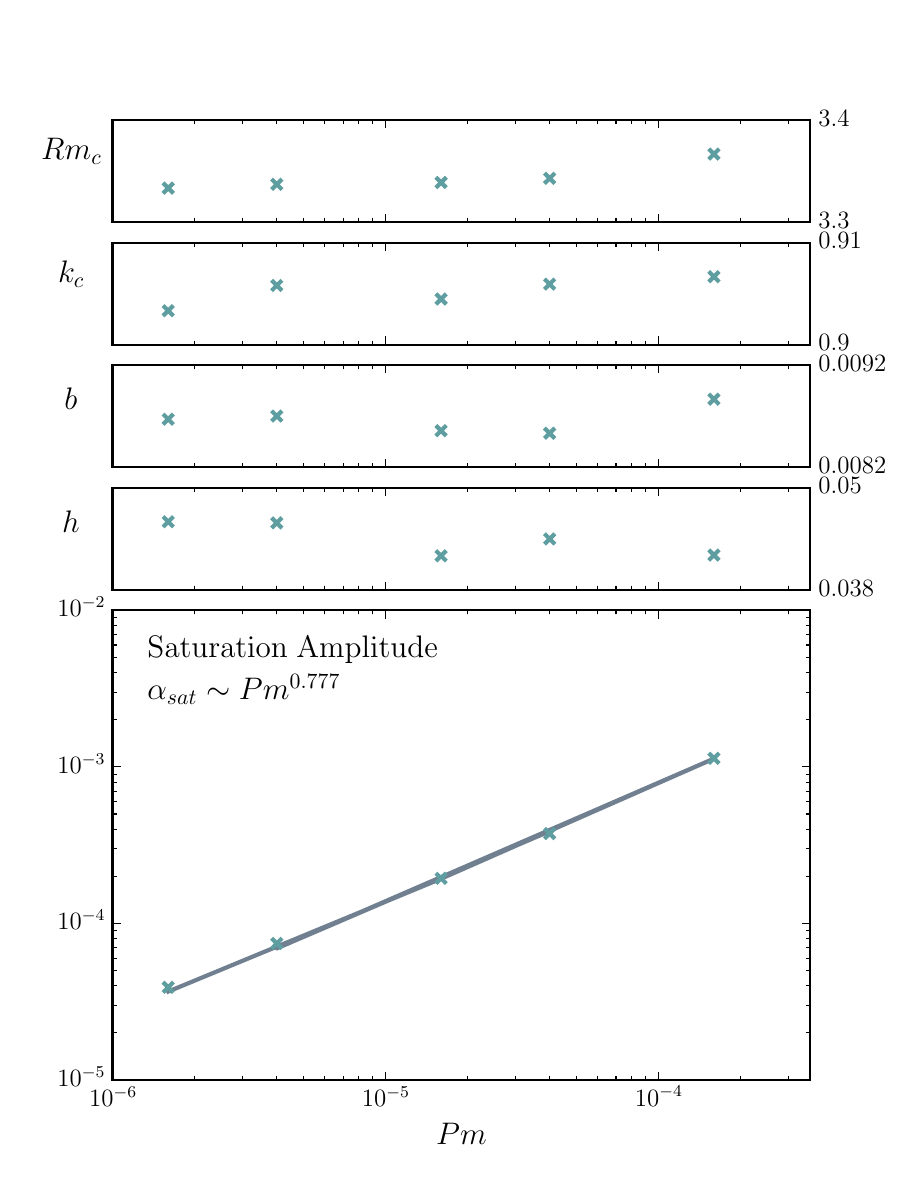}
\caption{Critical parameters $\reym_c$ and $k_c$, and coefficients of the Ginzburg-Landau equation (Equation \ref{eq:gle}) as a function of $\Pm$. Note the very weak dependence of the linear ($b$) and diffusive ($h$) coefficients on $\Pm$. The saturation amplitude $\alpha_{sat} = \sqrt{b/c}$ of the standard MRI system has a power law dependence on $\Pm$ which we measure to be $\alpha_{sat} \sim \Pm^{0.777}$. This scaling is driven by the $\Pm$ dependence of the nonlinear coefficient $c$.}\label{fig:coefficients}
\end{figure}

We find the marginal system as a function of the dimensionless parameters. The marginal stability curve for our standard MRI system is a hyperplane in $(\reym, \Pm, \Co, \Omega_2/\Omega_1, \mathrm{R}_1/\mathrm{R}_2$), but we hold all of these constant except for $\reym$. To analyze the MRI system at marginality, we fix the parameters listed in Table \ref{table:parameters} and determine the critical $\reym$ and vertical wavenumber $k_z$ by repeatedly solving the linear MRI system to determine the smallest parameter values for which the fastest growing mode has zero growth rate. That is, we solve the linear eigenvalue problem for eigenvalues $\sigma = \gamma + i \omega$ and determine the parameters which yield $\gamma = 0$. Figure \ref{fig:growth_rates} shows linear MRI growth rates $\gamma$ in the $(\reym, k_z)$ plane. For the fiducial standard MRI parameters in Table \ref{table:parameters} we find critical parameters $\reym_c = 3.30$ and $k_c = 0.901$.

\begin{figure*}
\centering
\includegraphics[width=\textwidth]{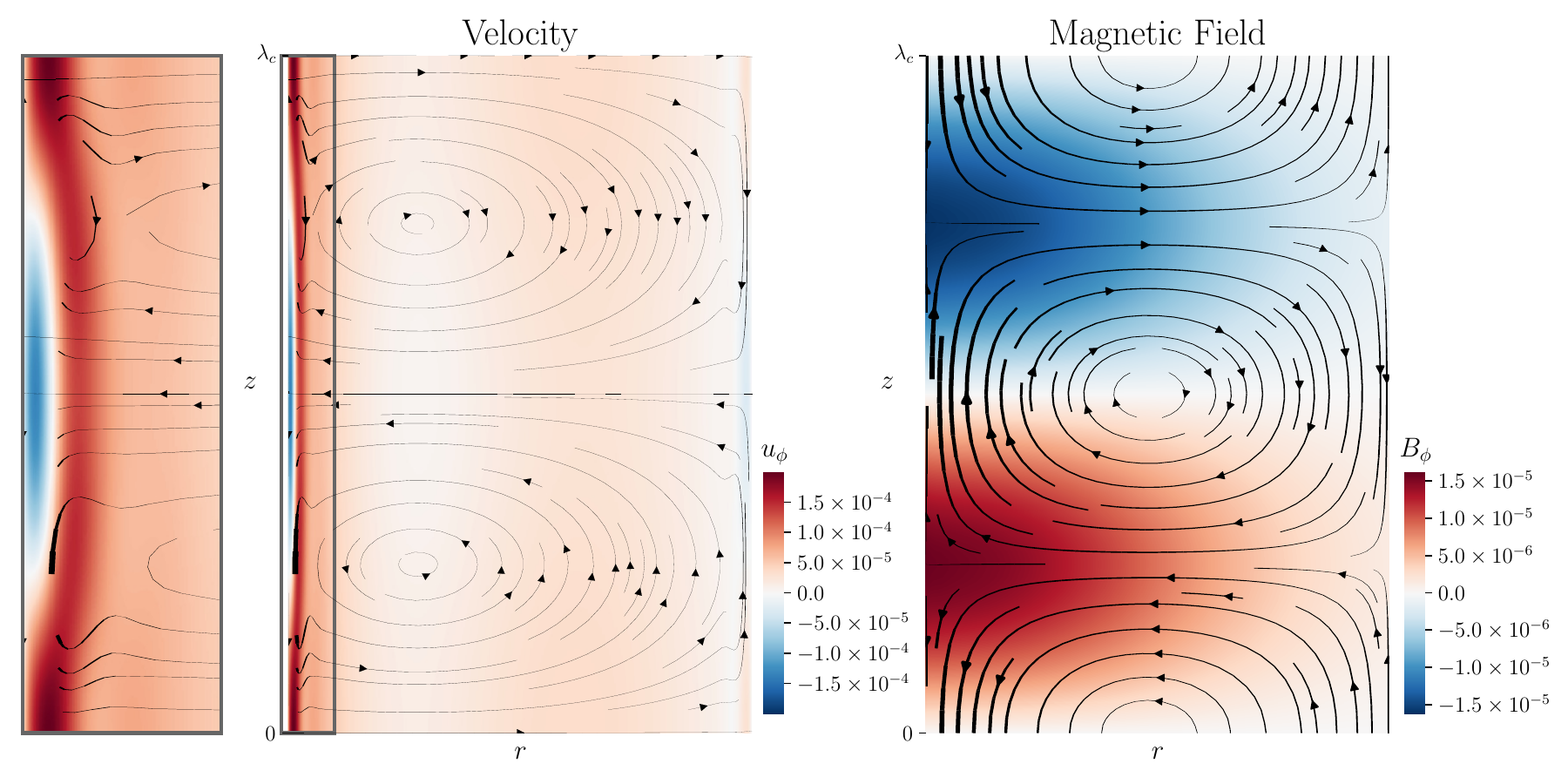}
\caption{Perturbation structure for the velocity and magnetic field of the fiducial standard MRI case, including first and second order perturbations. Leftmost panel is a radially zoomed-in section of the velocity perturbation structure, to better show the boundary layer-driven structure at the inner cylinder. Colors are azimuthal velocity and magnetic field perturbations, and streamfunctions show the perturbation structure in the $r, z$ plane. The width of the streamfunctions is proportional to the speed and magnetic field strength in the $r, z$ plane for the velocity and magnetic field, respectively. Vertical domain covers one critical wavelength $\lambda_c = 2\pi / k_c$. We use the constant saturation amplitude $\alpha_{s} = 3.9 \times 10^{-5}$ derived for this case, and a small parameter $\epsilon = 0.5$.}\label{fig:smri_streamfuncs}
\end{figure*}

Just as in the weakly nonlinear analyses of \citet{Umurhan:2007hs} and Paper I, we tune the system away from marginality by taking $B_0 \rightarrow B_0\left(1 - \epsilon^2\right)$, where the small parameter $\epsilon \ll 1$. We parameterize scale separation as $Z = \epsilon z$ and $T = \epsilon^2 t$, where $Z$ and $T$ are slowly varying spatial and temporal scales, respectively. We group the fluid variables into a state vector $\mathbf{V} = \left[\Psi, u, A, B\right]^{\mathrm{T}}$, such that the full nonlinear system in Equations \ref{eq:Psi_perturbed} - \ref{eq:nonlinear_B} can be expressed as

\beq\label{eq:unperturbed_matrix_eqns}
\mathcal{D}\partial_t\mathbf{V} + \mathcal{L}\mathbf{V} + \epsilon^2 \widetilde{\mathcal{G}} \mathbf{V} + \xi \widetilde{\mathcal{H}} \mathbf{V} +  \mathbf{N} = 0,
\eeq

where $\mathcal{D}$, $\mathcal{L}$, and $\widetilde{\mathcal{G}}$ are matrices defined in Appendix \ref{app:basic_equations}, and $\mathbf{N}$ is a vector containing all nonlinear terms. We expand the variables in a perturbation series 

\beq
\mathbf{V} = \epsilon \mathbf{V}_1 + \epsilon^2 \mathbf{V}_2 + \epsilon^3 \mathbf{V}_3 + h.o.t.
\eeq

 The perturbed system can then be expressed at each order by the equations

\begin{align}
\mathcal{O}(\epsilon)&: \mathcal{L} \mathbf{V}_1 + \xi \widetilde{\mathcal{H}} \mathbf{V}_1 + \mathcal{D} \partial_t \mathbf{V}_1 = 0 \label{eq:ordere}\\
\mathcal{O}(\epsilon^2)&: \mathcal{L} \mathbf{V}_2 + \xi \widetilde{\mathcal{H}} \mathbf{V}_2 + \mathcal{D} \partial_t \mathbf{V}_2 + \widetilde{\mathcal{L}}_1 \partial_Z \mathbf{V}_1 \nonumber\\
& + \xi \mathcal{H} \partial_Z \mathbf{V}_1 + \mathbf{N}_2 = 0 \label{eq:ordere2}\\
\mathcal{O}(\epsilon^3)&: \mathcal{L}\mathbf{V}_3 + \xi \widetilde{\mathcal{H}} \mathbf{V}_3 + \mathcal{D} \partial_t \mathbf{V}_3 + \mathcal{D} \partial_T \mathbf{V}_1 \nonumber\\
&+ \widetilde{\mathcal{L}}_1 \partial_Z \mathbf{V}_2 + \xi \mathcal{H}\partial_Z \mathbf{V}_2 + \widetilde{\mathcal{L}}_2 \partial_Z^2 \mathbf{V}_1 - \xi \widetilde{\mathcal{H}} \mathbf{V}_1 \nonumber\\
& + \widetilde{\mathcal{G}} \mathbf{V}_1 + \mathbf{N}_3 = 0.\label{eq:ordere3}
\end{align}

The nonlinear terms $\mathbf{N}_2$ and $\mathbf{N}_3$ which appear at $\mathcal{O}(\epsilon^2)$ and $\mathcal{O}(\epsilon^3)$, respectively, contain the nonlinear interaction between MRI modes. The system is weakly nonlinear because this mode interaction occurs in a controlled way. At $\mathcal{O}(\epsilon^2)$, the nonlinear terms represent the interaction of linear ($\mathcal{O}(\epsilon)$) MRI modes with themselves and their complex conjugates. At $\mathcal{O}(\epsilon^3)$, the nonlinear terms contain the interaction between first- and second-order MRI modes. See Appendix \ref{app:basic_equations} for the definition of matrices and a thorough derivation, and Appendix \ref{app:nonlinear_terms} for the detailed form of the nonlinear vectors. We emphasize that Equations \ref{eq:ordere} - \ref{eq:ordere3} have the same form as these equations in the narrow gap case, although the matrices, which contain all radial derivatives, are significantly different in this wide gap formulation. This is because we do not have slow variation in the radial dimension. In the standard MRI case, $\sigma = 0$ at marginality and so the $\partial_t$ terms drop out of the equations. For the helical MRI case, however, $\sigma$ has a nonzero imaginary component even at threshold, so we must formally include these terms in our perturbation expansion. The slow variation in $Z$ and $T$ are parameterized as an amplitude function $\alpha(Z, T)$ which modulates the flow in these dimensions. This parameterization coupled with the boundary conditions lead us to an ansatz linear solution 

\beq
\label{eq:ansatz}
\mathbf{V}_1 = \alpha(Z, T) \mathbb{V}_{11}(r) e^{i k_z z + \sigma t} + c.c.,
\eeq

where the radial variation is contained in $\mathbb{V}_{11}$, and $\sigma = \gamma + i\omega$. 

We solve the equations at each order using Dedalus, an open source pseudospectral code. We solve the radial portion of the eigenvectors on a basis of Chebyshev polynomials subject to our radial boundary conditions. We use a 512-component Chebyshev grid, and confirm numerical convergence at $1.5\times$ the resolution. This is sufficient to determine convergence because of the faster-than-exponential convergence of spectral methods \citep{Boyd:2001aa}.
We solve Equation \ref{eq:ordere} as a linear eigenvalue problem, and Equation \ref{eq:ordere2} as a linear boundary value problem. 
The result of the weakly nonlinear analysis is a single amplitude equation for $\alpha$. This amplitude equation is found by enforcing a solvability condition on Equation \ref{eq:ordere3}. We find

\beq
 \label{eq:gle}
\partial_T \alpha = b \alpha + d \partial_Z^2 \alpha - c \alpha \left|\alpha^2\right|,
\eeq

a Ginzburg-Landau equation (GLE). The GLE governs the weakly nonlinear amplitude behavior in a wide range of physical systems, including the narrow gap MRI \citep{Umurhan:2007hs}, Rayleigh-B\'enard convection \citep{Newell:1969wr}, and hydrodynamic TC flow \citep[e.g.][]{Recktenwald:1993}. We emphasize that this is a model equation, valid only near marginality \citep{Cross:1993el}. The dynamics of the GLE are determined by its coefficients, which are in turn determined by the linear eigenfunctions and nonlinear vectors plotted in Figures \ref{fig:linear_eigenfunctions} and \ref{fig:N2_N3}. Equation \ref{eq:gle} contains three coefficients: $b$, which determines the linear growth rate of the system, $d$, a diffusion coefficient, and $c$, the coefficient of the nonlinear term. When all of the coefficients of Equation \ref{eq:gle} are real, this is known as the real GLE, although the amplitude $\alpha$ is in general complex. The real GLE is subject to several well-studied instabilities, including the Ekhaus and Zig-Zag instabilities. When the coefficients are complex, we have the complex GLE, a source of even richer phase dynamics than its real counterpart \citep[see][]{Aranson:2002} for a thorough review.

\section{Results}
\label{sec:results}

\begin{figure}
\centering
\includegraphics[width=\columnwidth]{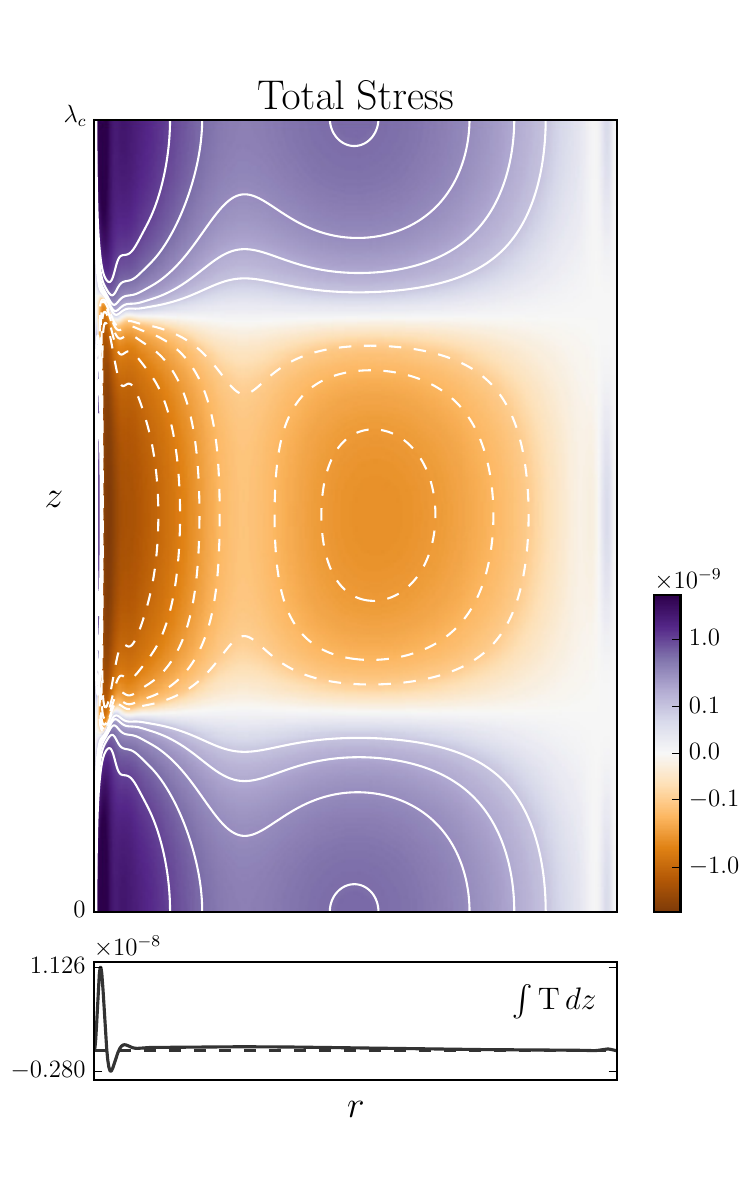}
\caption{Total stress $\mathrm{T}(r, z) = u_r u_\phi - \Co B_r B_\phi$, i.e. the sum of the Reynolds and Maxwell stresses for the fiducial standard MRI parameters (top panel). Bottom panel shows the vertically integrated stress $\int \mathrm{T}(r, z) \, dz$.}\label{fig:total_stress}
\end{figure}

\subsection{Standard MRI}
\label{sec:SMRI}
For the standard MRI we derive a real GLE. Here we note a departure from the behavior of the narrow gap system. The purely conducting boundary condition states that the axial component of the current ($\mathbf{J}_z = [\nabla \times \mathbf{B}]_z)$ must be zero at the walls. In the thin gap geometry, the purely conducting boundary condition on the azimuthal magnetic field is $\partial_x(B_y) = 0$ for axisymmetric perturbations. A spatially constant azimuthal field satisfies both the thin-gap MRI equations and this boundary condition. This neutral mode is formally included in the analysis of \citet{Umurhan:2007hs} and yields a second amplitude equation in the form of a simple diffusion equation. This amplitude equation decouples from the GLE because of the translational symmetry of the thin-gap geometry. Because that symmetry is not preserved in the wide-gap case, \citet{Umurhan:2007hs} postulate that slow variation in the wide-gap geometry will be governed by two coupled amplitude equations. However, the purely geometric term in Equation \ref{eq:Bphi_perturbed} prevents the wide-gap geometry from sustaining a neutral mode. We note that a neutral mode of the form $B_\phi(r) \propto \frac{1}{r}$ would exist in a resistance-free approximation. Here, however, this mode does not exist and we derive a single real GLE as the amplitude equation of the standard MRI.

The preservation of symmetries in the thin-gap geometry is worth a closer look, as its absence in the wide gap case is the source of many differences in the systems. \citet{Latter:2015} point out that in the ideal limit ($\nu, \eta \rightarrow 0$), the linearized system described by the lefthand side of Equations \ref{eq:Psi_perturbed} - \ref{eq:Bphi_perturbed} can be expressed as a Schr{\"o}dinger equation for the radial velocity. Similarly combining equations to obtain a single expression for $\Psi$, we find that the thin-gap limit, linear, ideal MRI can be expressed as

\beq
\partial_x^2 \Psi + k_z^2 \mathrm{U}(x) \Psi = 0
\eeq 

where $\mathrm{U}(x) = {3}/{v_A^2 k_z^2} + 1$ at marginality. This form is not unique to the ideal MHD case, though the ideal approximation simplifies the expression considerably. When no-slip radial boundary conditions are applied, the thin-gap MRI system resembles a particle in a box with a radially constant potential well. Thus thin-gap linear MRI modes must be eigenstates of parity. These symmetries are preserved in the nonlinear MRI vectors because they are nonlinear combinations of lower-order eigenfunctions. In the wide gap case, the ``potential" $\mathrm{U}(r)$ varies with $r$, so symmetric and antisymmetric modes are no longer required. This lack of symmetry is readily apparent in the eigenfunctions and nonlinear vectors in Figures \ref{fig:linear_eigenfunctions} and \ref{fig:N2_N3}, both of which display enhanced boundary layer activity at the inner boundary as compared to the outer boundary. The inner and outer boundary layers are symmetric in the thin gap case (see Figure \ref{fig:thin_wide_comparison}).

The form of the nonlinear terms, detailed in Appendix \ref{app:nonlinear_terms}, represent a departure from the thin-gap theory. The narrow gap nonlinear terms at both second and third orders are linear combinations of Jacobians. The nonlinear terms in the wide-gap case differ from their thin-gap analogues with the addition of vertical advective terms. These terms derive from the advective derivatives in the momentum and induction equations, but are filtered out in the thin-gap approximation. The nonlinear terms ultimately determine the saturation amplitude of the system, as described below.

\begin{figure*}
\centering
\includegraphics[width=\textwidth]{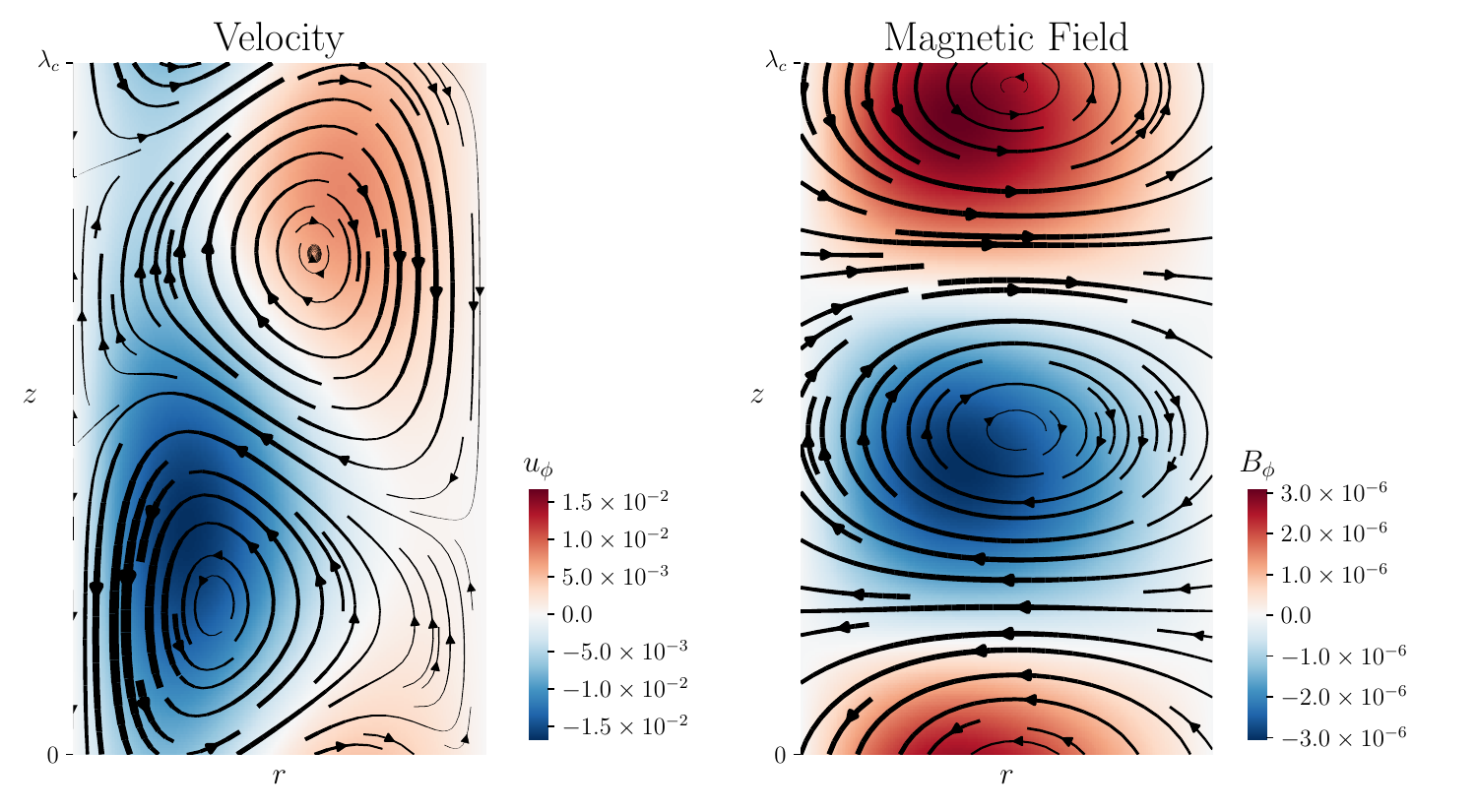}
\caption{As in Figure \ref{fig:smri_streamfuncs} but for the fiducial helical MRI case, including first and second order perturbations.}
\end{figure*}

We examine the behavior of the wide gap MRI system as a function of $\Pm$ in the regime $\Pm \ll 1$. Figure \ref{fig:coefficients} shows the critical parameters $k_c$ and $\reym$ as a function of $\Pm$, as well as the GLE linear coefficient $b$ and the diffusion coefficient $d$. From Equation \ref{eq:gle} it is apparent that the asymptotic saturation amplitude is $\alpha_{s} = \pm \sqrt{b/c}$, and we plot the dependence of $\alpha_{s}$ on $\Pm$ in the bottom panel of Figure \ref{fig:coefficients}. 
Note that because $\reym$ is essentially constant as a function of $\Pm$, the saturation amplitude is equivalently sensitive to $\reye^{-1}$. We find by fitting the data that the saturation amplitude scales as $\alpha_{s} \sim \Pm^{0.777}$. For these same boundary conditions, \citeauthor{Umurhan:2007hs} (\citeyear{Umurhan:2007hs}) find that the narrow gap saturation amplitude scales as $\Pm^{2/3}$. They find that this amplitude dependence is driven by the $\Pm^{1/3}$ dependence of the linear boundary layer. Boundary layer analysis similarly reveals a $\nu^{1/3}$ dependence for the radial extent of the boundary layer in TC flow \citep{Goodman:2002ix}.
Figure \ref{fig:thin_wide_comparison} shows the structure of the third-order nonlinear term $N_{31}^{(A)}$ as a function of $\Pm$ for both the narrow and wide gap standard MRI. $N_{31}$ is the vector that determines the GLE coefficient $c$, and thus the scaling of the saturation amplitude because of the insensitivity of $b$ to $\Pm$ (see Appendix \ref{app:basic_equations} for the wide gap case, and \citet{Umurhan:2007hs} and Paper I for the narrow gap equations). Clearly, the width of the boundary layers scales with $\Pm$ in both the wide and narrow gap MRI. This translates to a steeper saturation amplitude $\Pm$ dependence in the wide gap case.

Because it is governed by a real GLE, the saturated standard MRI state may be unstable to the Eckhaus instability, in which the wavelength of the large-scale pattern is adjusted \citep[e.g.][]{Hoyle:2006}. Preliminary investigation of the GLE behavior for the standard MRI coefficients derived here indicates that when the simulated vertical domain is large (i.e. spans multiple critical wavelengths), the amplitude function is modulated in $Z$, but always be bounded by $\alpha_{s} = \pm \sqrt{b/c}$, as must be the case for the one-dimensional real GLE. In Figure \ref{fig:smri_streamfuncs} we plot the saturated state perturbation structure of the fiducial standard MRI, up to and including second order disturbances. We use a constant saturation amplitude, but note that a nonconstant $\alpha_{s}$ would introduce more vertical structure. We similarly plot the total stress, i.e. the sum of the Reynolds and Maxwell stresses in our domain (Figure \ref{fig:total_stress}). As in Paper I, we find that the saturation mechanism for weakly nonlinear TC flow is a combination of reduced shear and redistributed and amplified background $B_z$. This strongly suggests that the underlying physics remains the same in the wide gap geometry, despite the addition of curvature terms.

\subsection{Helical MRI}

When $\xi$ in Equation \ref{eq:unperturbed_matrix_eqns} is nonzero, the helical MRI arises. We examine a single fiducial helical MRI case, for the parameters used by \citet{Hollerbach:2005tr}, listed in Table \ref{table:parameters}. The helical MRI is an overstability, so the ansatz linear eigenvector we consider (Equation \ref{eq:ansatz}) is characterized by a complex temporal eigenvalue $\sigma$. For our fiducial parameters, the marginal mode has a frequency $\omega = 0.153$. This means that the helical MRI modes are traveling waves, moving in the $z$ direction with a phase velocity $\omega/k_c$.

At the conclusion of the weakly nonlinear analysis, we find that the coefficients of Equation \ref{eq:gle} are complex. The marginal helical MRI is thus described by a complex GLE. This difference in character between the amplitude equations that modulate the weakly nonlinear standard and helical MRI is a consequence of the same property that makes the helical MRI an overstability. With the introduction of an azimuthal component, the background magnetic field acquires a handedness that is not present in a purely axial field. The helical MRI eigenvectors are therefore free to be out of phase with one another. In our perturbation series, the helical MRI modes interact within and between orders with modes which carry different phases, leading to complex GLE coefficients.

The phase dynamics of the complex GLE are well-studied in a variety of systems, and depend on the values of the GLE coefficients. The complex GLE may be unstable to traveling wave instabilities such as the Benjamin-Feir instability, a generalization of the Ekhaus instability. The complex GLE can also admit spatiotemporal chaos, and various classes of coherent structures \citep{Aranson:2002}. Although a detailed description of the phase dynamics in the helical MRI is beyond the scope of this work, we note that such long-wavelength, long-timescale behavior may be observed in liquid metal helical MRI experiments. 

\section{Discussion}
\label{sec:discussion}
In this work we carry out a formal weakly nonlinear multiscale analysis of the MRI in a Taylor-Couette flow. We analyze both the standard and helical MRI, which differ only in the geometry of their imposed background magnetic fields. We find that the amplitude function, which governs the behavior of the system on long length- and timescales, obeys a real GLE for the case of the standard MRI, and a complex GLE for the helical MRI. These two systems are thus subject to different large-scale phase dynamics. 

Our work should be placed in the broader context of emergent pattern formation in physical systems. The real Ginzburg-Landau equation derived here governs the slow-parameter evolution of the standard MRI close to threshold. The GLE arises in a number of other physical systems, and in each case it is a consequence not of the particular physics at hand, but of the underlying symmetries in the problem. Here we make a phenomenological comparison to two other systems that give rise to a GLE. The first and perhaps most famous is Rayleigh-B\'enard convection, in which a fluid between two plates is heated from below \citep{Newell:1969wr}. If we take the plane of the fluid to be infinite in the horizontal plane, the system is initially translationally symmetric. 
At the onset of convection the system undergoes a symmetry breaking, forming rolls, or convection cells, which break the horizontal translational invariance. Analogously, the standard MRI system considered here is initially vertically translationally symmetric, because we idealize the TC device as an infinitely long cylinder. The MRI breaks this symmetry, forming cells along the vertical length of the domain. Just as Rayleigh-B\'enard cells transport heat vertically, the MRI cells transport angular momentum horizontally. The symmetry breaking of each of these systems is described near onset by the real GLE. 

A real GLE has also been found to describe the formation of zonal flows out of magnetized turbulence in a model system \citep{Parker:2013hy, Parker:inpress}. Zonal flows are axisymmetric structures, large-scale and long-lived, which form spontaneously out of turbulence. They have recently been observed in some numerical studies of the MRI, and have generated considerable interest for their possible role in planet formation in protoplanetary disks \citep{Johansen:2009uj,Kunz:2013}. The present work is of course an idealized geometry, and we make no attempt to model a realistic protoplanetary disk environment. However, it is worth noting that the GLE we derive implies that axisymmetric, large-scale, long-lived structures are a generic feature of the MRI in the weakly nonlinear regime. This work provides a mathematical description of the MRI as a pattern-forming process, but much remains to be understood, particularly involving the application of this model system to realistic astrophysical disks. Paper I establishes that the GLE will arise in the shearing box approximation in the presence of ambipolar diffusion, and this work demonstrates that the pattern-forming behavior is not an artifact of the local geometry. The stage is thus set to apply this theory to more astrophysical conditions in either the global geometry or a local approximation. Of course our current model is most directly relevant to TC flows, and we emphasize that laboratory MRI experiments stand poised to observe the MRI-driven pattern formation predicted here. 

We detail several avenues for future work, which highlight the application of this theory to both laboratory experiments and astrophysical disks:

\begin{itemize}
\item Our theory may be applied to a specific experimental apparatus to model the predicted saturated state. One can then ask whether GLE dynamics should be detectable, especially over endcap-driven flows.
\item The saturation properties of different rotation profiles may be compared by direct application of the theory developed here.
\item Vertical stratification may be added to the base state, constructing a more realistic model of global vertical disk structure. 
\item Other nonideal MHD effects such as the Hall effect and ambipolar diffusion are straightforward additions to this model, and are of particular interest for understanding protoplanetary disks. 
\item The background magnetic field geometry may be generalized to include radial variation, another feature relevant to astrophysical disks.
\item The radial boundary conditions considered here may be expanded to mimic astrophysical disks rather than TC devices.
\item Our theory can be compared to simulations in both the weakly and strongly nonlinear regimes: both the pattern selection at saturation and the $\Pm^{0.777}$ scaling can be directly tested. 
\end{itemize}

This is the first weakly nonlinear analysis of the MRI in a cylindrical geometry, and is thus the global analogue of similar analyses in local approximations \citep{Umurhan:2007hs,Vasil:2015}. Understanding the connection between local and global MRI modes is crucial for interpreting simulation results across domain geometries. (The relationship between local and global \textit{linear} MRI modes is investigated in \citeauthor{Latter:2015} \citeyear{Latter:2015}.) Phenomena such as saturation and the development of turbulence depend critically on the nature of the underlying MRI modes. The formalism presented here describes analytically the weakly nonlinear behavior of global MRI modes. This treatment should be expanded to encompass more astrophysically relevant conditions, so that our understanding of complicated MRI phenomena may continue to make contact with analytical theory. 

\section{Acknowledgments}
S.E.C. was supported by a National Science Foundation Graduate Research Fellowship under grant No. DGE-16-44869. J.S.O. acknowledges support from NASA grant NNX16AC92G. We thank the anonymous referee for many thoughtful comments on both this and Paper I. We also thank Mordecai Mac Low, Jeremy Goodman, John Krommes, Geoff Vasil, and Ellen Zweibel for useful discussion.
\software{Dedalus: http://dedalus-project.org/}

%************* APPENDICES ************************%

\clearpage
\appendix

\section{Detailed Equations}\label{app:basic_equations}

Here we detail the perturbation analysis described in Section \ref{sec:wnl_analysis}. The perturbation series is described by Equations \ref{eq:ordere} - \ref{eq:ordere3}, where 

\beq
\mathcal{L} = \mathcal{L}_0 + \mathcal{L}_1 \partial_z + \mathcal{L}_2 \partial_z^2 + \mathcal{L}_3 \partial_z^3 + \mathcal{L}_4 \partial_z^4,
\eeq

\beq
\widetilde{\mathcal{L}}_1 =  \mathcal{L}_1 + 2\mathcal{L}_2\partial_z + 3\mathcal{L}_3\partial_z^2 + 4\mathcal{L}_4\partial_z^3
\eeq

\beq
\widetilde{\mathcal{L}}_2 = \mathcal{L}_2 + 3\mathcal{L}_3\partial_z + 6\mathcal{L}_4\partial_z^2
\eeq

\beq
\widetilde{\mathcal{G}} = \mathcal{G} \partial_z + \mathcal{L}_3 \partial_z^3,
\eeq

\beq
\widetilde{\mathcal{H}} = \mathcal{H} \partial_z,
\eeq

and the constituent matrices are defined as 

\beq
\mathcal{L}_0 = \left[\begin{matrix}
-\frac{1}{\reye} (-\frac{3}{r^4} \partial_r + \frac{3}{r^3}\partial_r^2 - \frac{2}{r^2}\partial_r^3 + \frac{1}{r}\partial_r^4) & 0 & 0 & 0 \\
0 & -\frac{1}{\reye} (\partial_r^2 + \frac{1}{r}\partial_r - \frac{1}{r^2}) & 0 & 0 \\
0 & 0 & -\frac{1}{\reym} (\partial_r^2 - \frac{1}{r} \partial_r) & 0 \\
0 & 0 & 0 & -\frac{1}{\reym} (\partial_r^2 + \frac{1}{r}\partial_r - \frac{1}{r^2}) \\
\end{matrix}\right]
\eeq

\beq
\mathcal{L}_1 = \left[\begin{matrix}
0 & -\frac{2}{r} u_0 & \Co (\frac{1}{r^2} \partial_r - \frac{1}{r}\partial_r^2) & 0 \\
\frac{1}{r^2} u_0 + \frac{1}{r}\partial_r u_0 & 0 & 0 & -\Co \\
-1 & 0 & 0 & 0 \\
0 & -1 & \frac{1}{r^2} u_0 - \frac{1}{r} \partial_r u_0 & 0 \\
\end{matrix}\right]
\eeq

\beq
\mathcal{L}_2 = \left[\begin{matrix}
-\frac{1}{\reye} (-\frac{2}{r^2}\partial_r + \frac{2}{r}\partial_r^2) & 0 & 0 & 0 \\
0 & -\frac{1}{\reye} & 0 & 0 \\
0 & 0 & -\frac{1}{\reym} & 0 \\
0 & 0 & 0 & -\frac{1}{\reym} \\
\end{matrix}\right]
\eeq

\beq
\mathcal{L}_3 = \left[\begin{matrix}
0 & 0 & -\Co \frac{1}{r} & 0 \\
0 & 0 & 0 & 0 \\
0 & 0 & 0 & 0 \\
0 & 0 & 0 & 0 \\
\end{matrix}\right]
\eeq

\beq
\mathcal{L}_4 = \left[\begin{matrix}
-\frac{1}{\reye}\frac{1}{r} & 0 & 0 & 0 \\
0 & 0 & 0 & 0 \\
0 & 0 & 0 & 0 \\
0 & 0 & 0 & 0 \\
\end{matrix}\right]
\eeq

\beq
\mathcal{G} = \left[\begin{matrix}
0 & 0 & \Co (\frac{1}{r^2}\partial_r - \frac{1}{r}\partial_r^2) & 0 \\
0 & 0 & 0 & -\Co \\
-1 & 0 & 0 & 0 \\
0 & -1 & 0 & 0 \\
\end{matrix}\right]
\eeq

\beq
\mathcal{H} = \left[\begin{matrix}
0 & 0 & 0 & \Co \frac{2}{r^2} \\
0 & 0 & 0 & 0 \\
0 & 0 & 0 & 0 \\
-\frac{2}{r^3} & 0 & 0 & 0 \\
\end{matrix}\right]
\eeq

\beq
\mathcal{D} = \left[\begin{matrix}
\frac{1}{r}\partial_r^2 + \frac{1}{r}\partial_z^2 - \frac{1}{r^2}\partial_r & 0 & 0 & 0 \\
0 & 1 & 0 & 0 \\
0 & 0 & 1 & 0 \\
0 & 0 & 0 & 1 \\
\end{matrix}\right]
\eeq

We solve the $\mathcal{O}(\epsilon)$ (linear) system, followed by the $\mathcal{O}(\epsilon^2)$ system in Equation \ref{eq:ordere2}. At second order in $\epsilon$, nonlinear terms arise which are formed by the interaction of first-order MRI modes with themselves and their complex conjugates. This mode interaction means that the second-order nonlinear term is 

\beq
\mathbf{N}_2 = |\alpha|^2 \mathbf{N}_{20} + \alpha^2 \mathbf{N}_{22} e^{2 i k_c z},
\eeq

where terms are grouped by z-dependence. See Appendix \ref{app:nonlinear_terms} for the full form of the nonlinear terms. Equation \ref{eq:ordere2} must therefore be solved as three separate systems of equations, one for each possible $z$ resonance: 

\begin{align}
\mathcal{L}\mathbf{V}_{20} + \xi \partial_z \mathcal{H} \mathbf{V}_{20} & = \mathbf{N}_{20}\\
\mathcal{L}\mathbf{V}_{21} + \xi \partial_z \mathcal{H} \mathbf{V}_{21} & = - \widetilde{\mathcal{L}}_1 \partial_Z \mathbf{V}_{11} - \xi \partial_Z \mathcal{H} \mathbf{V}_{11} \\
\mathcal{L}\mathbf{V}_{22} + \xi \partial_z \mathcal{H} \mathbf{V}_{22} & = \mathbf{N}_{22}
\end{align}

To find a bounded solution at $\mathcal{O}(\epsilon^3)$ we must eliminate secular terms: terms which are resonant with the solution to the linear homogenous equation $(\mathcal{L} + \xi \widetilde{\mathcal{H}}) \mathbf{V} = 0$ and cause the solution to grow without bound. Secular terms in our system are those that are resonant with the linear ansatz (Equation \ref{eq:ansatz}), i.e. terms with $e^{i k_c z}$ z-dependence. To eliminate these terms we enforce a solvability condition, which arises from a corollary to the Fredholm alternative. The Fredholm alternative states that if we consider a system of equations $\mathcal{L} \mathbf{V} = \mathbf{b}$ and its adjoint homogenous system $\mathcal{L}^\dagger \mathbf{V}^\dagger = 0$, only one of two conditions holds. Either there exists one and only one solution to the nonhomogenous system, or the homogenous adjoint equation has a nontrivial solution. The relevant corollary arises as a consequence of the second condition: if $\mathcal{L}^\dagger \mathbf{V}^\dagger = 0$ has a nontrivial solution, then $\mathcal{L} \mathbf{V} = \mathbf{b}$ has a solution if and only if $\langle \mathbf{V}^\dagger | b \rangle = 0$. 

We define the adjoint operator $\mathcal{L}^\dagger$ and solution $\mathbf{V}^\dagger$ as 

\beq
\langle \mathbf{V^\dagger} | (\mathcal{L} + \xi \widetilde{\mathcal{H}}) \mathbf{V} \rangle = \langle (\mathcal{L}^\dagger + \xi \widetilde{\mathcal{H}}^\dagger) \mathbf{V}^\dagger | \mathbf{V} \rangle,
\eeq

where the inner product is defined as 

\beq\label{eq:inner_product_def}
\langle \mathbf{V^\dagger} | \mathcal{L} \mathbf{V} \rangle = \frac{k_c}{2\pi} \int_{-\pi/k_c}^{\pi/k_c} \int_{r_1}^{r_2} \mathbf{V}^{\dagger*} \cdot \mathcal{L} \mathbf{V} \, r \mathrm{d} r \mathrm{d} z
\eeq

We derive the adjoint operator by successive integration by parts, to find 

\beq
\mathcal{L}^\dagger = \mathcal{L}_0^\dagger - \partial_z \mathcal{L}_1^\dagger + d_z^2 \mathcal{L}_2^\dagger - \partial_z^3 \mathcal{L}_3^\dagger + \partial_z^4 L_4^\dagger
\eeq

and 

\beq
\mathcal{H}^\dagger = - d_z \mathcal{H}^\mathrm{T},
\eeq

where 

\beq
\mathcal{L}_0^\dagger = \left[\begin{matrix}
-\frac{1}{\reye} (-\frac{3}{r^5} + \frac{3}{r^4}\partial_r - \frac{3}{r^3} \partial_r^2 + \frac{2}{r^2}\partial_r^3 + \frac{1}{r}\partial_r^4 ) & 0 & 0 & 0 \\
0 & -\frac{1}{\reye} (\frac{1}{r}\partial_r + \partial_r^2 - \frac{1}{r^2}) & 0 & 0 \\
0 & 0 &-\frac{1}{\reym} (\frac{3}{r}\partial_r + \partial_r^2) & 0 \\
0 & 0 & 0 & -\frac{1}{\reym} (\frac{1}{r}\partial_r + \partial_r^2 - \frac{1}{r^2}) \\
\end{matrix}\right],
\eeq

\beq
\mathcal{L}_1^\dagger = \left[\begin{matrix}
0 & \frac{1}{r^2} u_0 + \frac{1}{r}\partial_r u_0 & -1 & 0 \\
-\frac{2}{r}u_0 & 0 & 0 & -1 \\
\Co (\frac{1}{r^3} - \frac{1}{r^2}\partial_r - \frac{1}{r}\partial_r^2) & 0 & 0 & \frac{1}{r^2}u_0 - \frac{1}{r}\partial_r u_0 \\
0 & -\Co & 0 & 0 \\
\end{matrix}\right],
\eeq

\beq
\mathcal{L}_2^\dagger = \left[\begin{matrix}
-\frac{1}{\reye}(-\frac{2}{r^3} + \frac{2}{r^2}\partial_r + \frac{2}{r}\partial_r^2) & 0 & 0 & 0 \\
0 & -\frac{1}{\reye} & 0 & 0 \\
0 & 0 & -\frac{1}{\reym} & 0 \\
0 & 0 & 0 & -\frac{1}{\reym} \\
\end{matrix}\right],
\eeq

and $\mathcal{L}_3^\dagger = \mathcal{L}_3^\mathrm{T}$, $\mathcal{L}_4^\dagger = \mathcal{L}_4^\mathrm{T}$. The adjoint boundary conditions are selected to satisfy Equation \ref{eq:inner_product_def}, and differ depending on the boundary conditions enforced on the homogenous system. Specifically, the boundary conditions arise from the requirement that the integrands in Equation \ref{eq:inner_product_def} are zero at $r_1$ and $r_2$. For the conducting boundary conditions we apply to the standard MRI, the adjoint equation 

\beq\label{eq:adjoint}
(\mathcal{L}^\dagger + \xi \widetilde{\mathcal{H}}^\dagger)\mathbf{V}^\dagger = 0
\eeq

must be solved subject to the boundary conditions

\beq
\Psi^\dagger = \partial_r \Psi^\dagger = u^\dagger = A^\dagger = \partial_r (r B^\dagger) = 0.
\eeq

For the insulating case, the adjoint boundary conditions are

\beq
k \frac{I_0 (k r)}{I_1 (k r)} r A^\dagger - 2 A^\dagger - r \partial_r A^\dagger = 0 \, \, \mathrm{at} \, \,   r = r_1
\eeq

\beq
- k \frac{K_0 (k r)}{K_1 (k r)} r A^\dagger - 2 A^\dagger - r \partial_r A^\dagger = 0 \, \, \mathrm{at} \, \,   r = r_2
\eeq

We take the inner product of the adjoint homogenous solution with the terms in Equation \ref{eq:ordere3} that are resonant with $e^{i k_c z}$. This gives us

\beq
\langle \mathbb{V}^\dagger | \mathcal{D} \mathbb{V}_{11} \rangle \partial_T \alpha + \langle  \mathbb{V}^\dagger | \widetilde{\mathcal{G}} \mathbb{V}_{11} + \xi \widetilde{\mathcal{H}} \mathbb{V}_{11} \rangle \alpha +  \langle \mathbb{V}^\dagger | \widetilde{\mathcal{L}}_1 \mathbb{V}_{21} + \widetilde{\mathcal{L}}_2 \mathbb{V}_{11} + \xi \mathcal{H} \mathbb{V}_{21} \rangle \partial_Z^2 \alpha = \langle \mathbb{V}^\dagger | \mathbf{N}_{31} \rangle \alpha |\alpha|^2,
\eeq

or Equation \ref{eq:gle}, the Ginzburg-Landau Equation, where the coefficients are 

\beq
b = \langle  \mathbb{V}^\dagger | \widetilde{\mathcal{G}} \mathbb{V}_{11} + \xi \widetilde{\mathcal{H}} \mathbb{V}_{11} \rangle / \langle \mathbb{V}^\dagger | \mathcal{D} \mathbb{V}_{11} \rangle,
\eeq

\beq
h = \langle \mathbb{V}^\dagger | \widetilde{\mathcal{L}}_1 \mathbb{V}_{21} + \widetilde{\mathcal{L}}_2 \mathbb{V}_{11} + \xi \mathcal{H} \mathbb{V}_{21} \rangle / \langle \mathbb{V}^\dagger | \mathcal{D} \mathbb{V}_{11} \rangle,
\eeq

and

\beq
c = \langle \mathbb{V}^\dagger | \mathbf{N}_{31} \rangle / \langle \mathbb{V}^\dagger | \mathcal{D} \mathbb{V}_{11} \rangle. 
\eeq

%%%% NONLINEAR TERMS %%%%%%%%%%%
\section{Nonlinear Terms}\label{app:nonlinear_terms}

Here we detail the perturbative expansion of the nonlinear vector $\mathbf{N}$ in Equation \ref{eq:unperturbed_matrix_eqns}. 

\beq
\mathbf{N} = \epsilon^2 \mathbf{N}_2 + \epsilon^3 \mathbf{N}_3
\eeq

\beq
N_2^{\Psi}  = -J(\Psi_1, \frac{1}{r^2} \nabla^2 \Psi_1) - J(\Psi_1, -\frac{2}{r^3}\partial_r\Psi_1)
+ \Co J (A_1, \frac{1}{r^2} \nabla^2 A_1) + \Co J(A_1, -\frac{2}{r^3} \partial_r A_1) + \frac{2}{r} u_1 \partial_z u_1 - \Co \frac{2}{r} B_1 \partial_z B_1
\eeq

\beq
\begin{split}
N_2^{u} = -\frac{1}{r} J\left(\Psi_1, u_1\right) + \frac{1}{r} \Co J\left(A_1, B_1\right) - \frac{1}{r^2} u_1 \partial_z \Psi_1 + \Co \frac{1}{r^2} B_1 \partial_z A_1
\end{split}
\eeq

\beq
N_2^A = \frac{1}{r} J\left(A_1, \Psi_1\right)
\eeq

\beq
N_2^B = \frac{1}{r} J\left(A_1, u_1\right) - \frac{1}{r} J\left(\Psi_1, B_1\right) + \frac{1}{r^2} B_1 \partial_z \Psi_1 - \frac{1}{r^2} u_1 \partial_z A_1
\eeq

\beq
\begin{split}
N_3^{\Psi} & = -J(\Psi_1, \frac{1}{r^2} \nabla^2 \Psi_2) - J(\Psi_2, \frac{1}{r^2} \nabla^2\Psi_1) - 2 J (\Psi_1, \frac{1}{r^2}\partial_Z\partial_z \Psi_1) - J(\Psi_1, -\frac{2}{r^3}\partial_r \Psi_2) - J(\Psi_2, -\frac{2}{r^3}\partial_r \Psi_1) \\
& - \widetilde{J}(\Psi_1, \frac{1}{r^2} \nabla^2 \Psi_1) - \widetilde{J} (\Psi_1, -\frac{2}{r^3}\partial_r \Psi_1) + \Co J(A_1, \frac{1}{r^2}\nabla^2 A_2) + \Co J(A_2, \frac{1}{r^2}\nabla^2 A_1) + 2\Co J(A_1, \frac{1}{r^2}\partial_Z\partial_z A_1) \\ 
& + \Co J(A_1, -\frac{2}{r^3} \partial_r A_2 ) + \Co J(A_2, -\frac{2}{r^3} \partial_r A_1) + \Co \widetilde{J} (A_1, \frac{1}{r^2} \nabla^2 A_1) + \Co \widetilde{J} (A_1, -\frac{2}{r^3} \partial_r A_1) \\
& + \frac{2}{r} u_1 \partial_z u_2 + \frac{2}{r} u_2 \partial_z u_1 + \frac{2}{r} u_1 \partial_Z u_1 - \Co \frac{2}{r} B_1\partial_z B_2 - \Co \frac{2}{r} B_2 \partial_z B_1 - \Co \frac{2}{r} B_1 \partial_Z B_1
\end{split}
\eeq

\beq
\begin{split}
N_3^u & = -\frac{1}{r}J\left(\Psi_1, u_2\right) - \frac{1}{r}J\left(\Psi_2, u_1\right) - \frac{1}{r}\widetilde{J} \left(\Psi_1, u_1\right) + \frac{1}{r}\Co J\left(A_1, B_2\right) + \frac{1}{r} \Co J\left(A_2, B_1\right) + \frac{1}{r}\Co \widetilde{J}\left(A_1, B_1\right) \\
& - \frac{1}{r^2} u_1\partial_z \Psi_2 - \frac{1}{r^2} u_2 \partial_z \Psi_1 - \frac{1}{r^2} u_1 \partial_Z \Psi_1 + \Co \frac{1}{r^2} B_1 \partial_z A_2 + \Co \frac{1}{r^2} B_2 \partial_z A_1 + \Co \frac{1}{r^2} B_1 \partial_Z A_1
\end{split}
\eeq

\beq
N_3^A = \frac{1}{r} J\left(A_1, \Psi_2\right) + \frac{1}{r}J\left(A_2, \Psi_1\right) + \frac{1}{r} \widetilde{J}\left(A_1, \Psi_1\right)
\eeq

\beq
\begin{split}
N_3^B & = \frac{1}{r} J\left(A_1, u_2\right) + \frac{1}{r} J\left(A_2, u_1\right) + \frac{1}{r}\widetilde{J}\left(A_1, u_1\right) + \frac{1}{r} J\left(B_1, \Psi_2\right) + \frac{1}{r} J\left(B_2, \Psi_1\right) + \frac{1}{r} \widetilde{J} \left(B_1, u_1\right) \\ 
& + \frac{1}{r^2} B_1\partial_z \Psi_2 + \frac{1}{r^2} B_2 \partial_z \Psi_1 + \frac{1}{r^2} B_1 \partial_Z \Psi_1 - \frac{1}{r^2} u_1 \partial_z A_2 - \frac{1}{r^2} u_2 \partial_z A_1 - \frac{1}{r^2} u_1 \partial_Z A_1
\end{split}
\eeq

\bibliographystyle{plain}

\end{document}